\documentclass[5p,twocolumn,times,number]{elsarticle}

\usepackage{amssymb}
\usepackage{amsmath}
\usepackage{physics}
\usepackage[section]{placeins}
\usepackage{subcaption}
\usepackage{orcidlink}
\usepackage{xspace}

\usepackage{lineno}

\journal{Nuclear Instruments and Methods A}
\newcommand{\gevc}{\ensuremath{{\mathrm{\,Ge\kern -0.1em V\!/}c}}\xspace}
\def\nutb       {\ensuremath{\nub_\tau}\xspace}
\def\nut        {\ensuremath{\nu_\tau}\xspace}
\def\numb       {\ensuremath{\nub_\mu}\xspace}
\def\nub        {\ensuremath{\overline{\nu}}\xspace}
\newcommand{\dedx}{\ensuremath{\mathrm{d}\hspace{-0.1em}E/\mathrm{d}x}\xspace}
\def\degrees{\ensuremath{^{\circ}}\xspace}
\def\KL    {\ensuremath{K^0_{\scriptscriptstyle L}}\xspace}

\begin{document}

\begin{frontmatter}



\title{Muon identification with Deep Neural Network in the Belle~II K-Long and Muon detector}

\author{Zihan~Wang\,\orcidlink{0000-0002-3536-4950}}
\ead{zihanwa@hep.phys.s.u-tokyo.ac.jp}

\author{Yo~Sato\,\orcidlink{0000-0003-3751-2803}}
\author{Akimasa~Ishikawa\,\orcidlink{0000-0002-3561-5633}}
\author{Yutaka~Ushiroda\,\orcidlink{0000-0003-3174-403X}}
\author{Kenta~Uno\,\orcidlink{0000-0002-2209-8198}}
\author{Kazutaka~Sumisawa\,\orcidlink{0000-0001-7003-7210}}
\author{Naveen~Kumar~Baghel\,\orcidlink{0009-0008-7806-4422}}
\author{Seema~Choudhury\,\orcidlink{0000-0001-9841-0216}}
\author{Giacomo~De~Pietro\,\orcidlink{0000-0001-8442-107X}}
\author{Christopher~Ketter\,\orcidlink{0000-0002-5161-9722}}
\author{Haruki~Kindo\,\orcidlink{0000-0002-6756-3591}}
\author{Tommy~Lam\,\orcidlink{0000-0001-9128-6806}}
\author{Frank~Meier\,\orcidlink{0000-0002-6088-0412}}
\author{Soeren~Prell\,\orcidlink{0000-0002-0195-8005}}

\begin{abstract}
Muon identification is crucial for elementary particle physics experiments. 
At the Belle~II experiment, muons and pions with momenta greater than 0.7\gevc are distinguished by their 
penetration ability through the $K_L$ and Muon (KLM) sub-detector, 
which is the outermost sub-detector of Belle~II.
In this paper, we first discuss the possible room for $\mu/\pi$ identification performance improvement and  
then present a new method based on Deep Neural Network (DNN). 
This DNN model utilizes the KLM hit pattern variables as the input and thus can digest the penetration information better than the current algorithm. 
We test the new method in simulation and find that the pion fake rate (specificity) is reduced from 4.1\% to 1.6\% at a muon efficiency (recall) of 90\%. 
\end{abstract}




\end{frontmatter}


\section{Introduction}

The Belle II~\cite{belle2} experiment 
is a key player in the measurement of flavor physics
at the intensity frontier. 
It makes use of the asymmetric 7~GeV electron 
and 4~GeV positron 
collision data provided by the SuperKEKB~\cite{superkekb} collider, 
located in KEK, Tsukuba (Japan). 
With the center-of-mass energy mainly
set to the $\Upsilon(4S)$ resonance, data samples with large amounts 
of $B$ mesons, $D$ mesons, and $\tau$ leptons are 
produced for various physics studies. 
Among them, the measurement of inclusive $b\to s \mu^+ \mu^-$ process is 
of special interest. 
It is a Flavor Changing Neutral Current (FCNC) decay that in SM proceeds through higher-order loop diagrams, which could be competitive with physics beyond the standard model amplitudes.  
In this measurement, one of the largest sources of peaking background comes 
from the $B\to X \pi^+ \pi^-$, whose branching 
fraction is three orders of magnitude larger than that of the signal process in the standard model. 
For this reason, typically a pion fake rate, defined as the probability that a pion is mis-identified as a muon, 
smaller than 2\% is required to ensure a good signal-to-noise ratio. 
In addition, muon identification also plays an important role in other physics topics, 
like lepton flavor universality 
test in $b\to c\tau^-\nutb$, where the $\tau$ lepton is reconstructed from the $\tau^- \to \mu^- \nut \numb$ decay.

In this work, we focus only on $\mu/\pi$ separation at the $K_L$ and Muon (KLM) sub-detector~\cite{klm}. 
Compared to muons, which only interact electromagnetically with detector materials, pions have strong interaction with iron, resulting in weaker penetration capability, higher probability of multiple-scattering, and larger cluster size if a hadronic shower is produced. 
Currently, muon identification in the KLM
is performed by a likelihood-based algorithm called muonID. 
To improve the muon identification performance, we develop a new algorithm based on Deep Neural Network (DNN), which uses the hit pattern as input. 

In the following, we will describe the Belle~II detector system 
for muon identification, 
introduce the muonID algorithm and discuss the possible room for performance improvement, 
and present the newly developed DNN based algorithm. 

\section{The Belle~II detector}

\begin{figure}[h]
    \centering
    \includegraphics[width = 0.48 \textwidth]{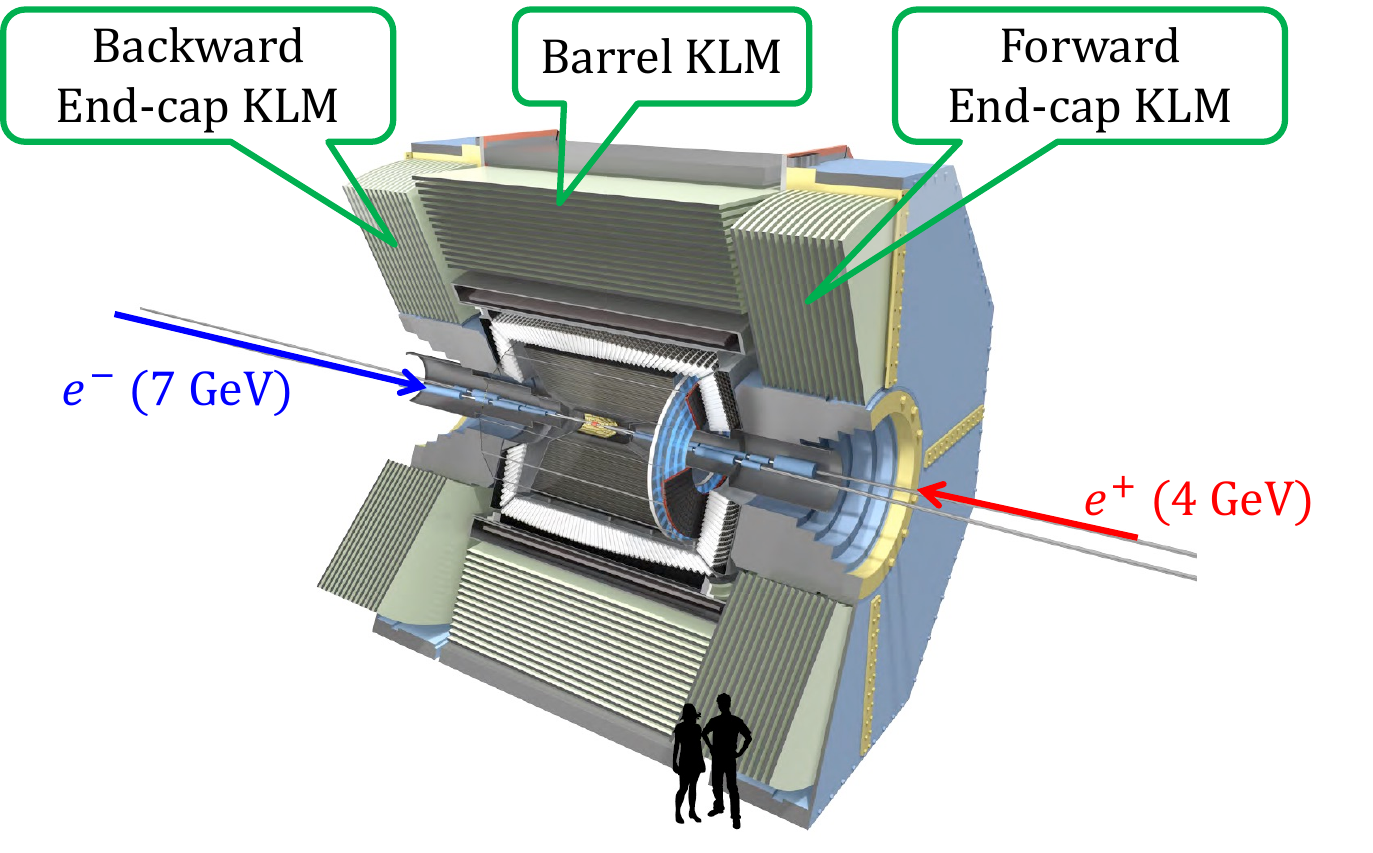}
    \caption{Belle~II detector. KLM sensor planes are placed in the gaps between the iron layers of the magnetic flux return. }
    \label{fig:klm}
\end{figure}

The Belle II detector shown in Fig.~\ref{fig:klm} 
has a cylindrical geometry and includes a
two-layer silicon-pixel detector~(PXD) surrounded by a four-layer
double-sided silicon-strip detector~(SVD)
and a 56-layer central drift chamber~(CDC). These detectors
reconstruct tracks of charged particles.  The
symmetry axis of these detectors, defined as the $z$ axis, is almost
coincident with the direction of the electron beam.  Surrounding the
CDC, which also provides \dedx energy loss measurements, is a
time-of-propagation counter~(TOP) in the
central region and an aerogel-based ring-imaging Cherenkov
counter~(ARICH) in the forward region.  These detectors provide
charged-particle identification.  Surrounding the TOP and ARICH is an
electromagnetic calorimeter~(ECL) based on CsI(Tl) crystals that
primarily provides energy and timing measurements for photons and
electrons. Outside of the ECL is a superconducting solenoid
magnet. Its flux return is instrumented with sensors to detect muons, 
$K^0_L$ mesons, and
neutrons. The solenoid magnet provides a 1.5~T magnetic field that is
oriented parallel to the $z$ axis. 

The KLM sub-detector consists of 4.7~cm thick iron plates alternated with 4.4~cm thick active layers. 
The octagonal barrel KLM (BKLM) is made of 14 iron plates and 15 detector layers, 
where the sensors in the inner two layers are plastic scintillator,  
and those in the outer 13 layers are Resistive Plate Chamber (RPC). 
The forward end-cap KLM (EKLM) consists of 14 iron plates and 14 detector layers of plastic scintillator, while there are 12 detector layers in the backward EKLM. 
The iron plates have a thickness equivalent to  more than 3.9 interaction lengths. 
Each detection layer is composed of two planes of strip sensors arranged orthogonally to give 2-dimensional coordinates. 
In KLM, one hit is defined as the overlap region of two strip clusters 
in the two perpendicular sensor planes. 
The KLM covers the polar angle range $20\degrees < \theta < 155\degrees$ 
with respect to the beam axis. Muons with a momentum above $0.7~\gevc$ penetrate the first layer of the KLM 
and the majority of muons traverse it completely if their momentum exceeds about $1.5\gevc$. 

\section{Likelihood-based muonID}

The traditional muonID algorithm consists of two major steps: 
(a) track extrapolation and hit association (assuming the muon hypothesis) to estimate the penetration path inside the KLM and (b) likelihood extraction 
based on the difference between extrapolation and observation.  

\subsection{Track extrapolation and hit association}
\label{sec:Track extrapolation}

The track extrapolation is performed by Geant4E~\cite{geant4}. 
Tracks reconstructed by PXD, SVD, and CDC are extrapolated to 
KLM with a muon hypothesis, considering characteristic energy loss \dedx and 
multiple scattering effects to estimate the track momentum and direction. 
Muons are assumed to not decay or interact through other physics processes.
After extrapolating to each KLM layer, the algorithm searches for a single hit associated with the track with a $\chi^2$ method.
The $\chi^2$ of each hit on the corresponding layer, 
which reflects the deviation of extrapolation position to the hit position, 
is calculated as 
\begin{equation}
    \label{eq:chi2}
    \chi^2 = \frac{(x_{ext} - x_{hit})^2}{\sigma_{ext}^2 + \sigma_{hit}^2},
\end{equation}
where $x_{hit(ext)}$ represents the hit (extrapolation) position, 
$\sigma_{ext}$ is the extrapolation uncertainty given by Geant4E, and 
$\sigma_{hit}$ is the hit position resolution summarized in~\cite{belle2}.  
This $\chi^2$ is calculated separately along the two directions of the strip sensors. 
The hit with the smallest sum of $\chi^2$ values in the two directions is selected as the hit associated with the track, if it also satisfies $\chi^2 < (3.5)^2$ in both directions.
If the associated hit exists, 
the extrapolated track properties are adjusted with respect to the hit using Kalman-filter~\cite{trackfinding}. 
Extrapolation stops when the particle's energy falls below 2~MeV or the track exits the detector. 
The number of layers crossed by extrapolation is denoted as $N_{ext}$. 

\subsection{Likelihood extraction}

Binary muonID is defined as the likelihood 
ratio of the muon and pion hypotheses: $\mathcal{L}_\mu/(\mathcal{L}_\mu + \mathcal{L}_\pi)$. 
In KLM, the likelihood with hypothesis $t$ ($\mu$ or $\pi$) 
is defined as the product of longitudinal and transverse likelihoods: $\mathcal{L}_t = \mathcal{L}^{\rm long}_t \times\mathcal{L}^{\rm trans}_t$. 

\begin{figure}[h]
    \centering
    \includegraphics[width = 0.3 \textwidth]{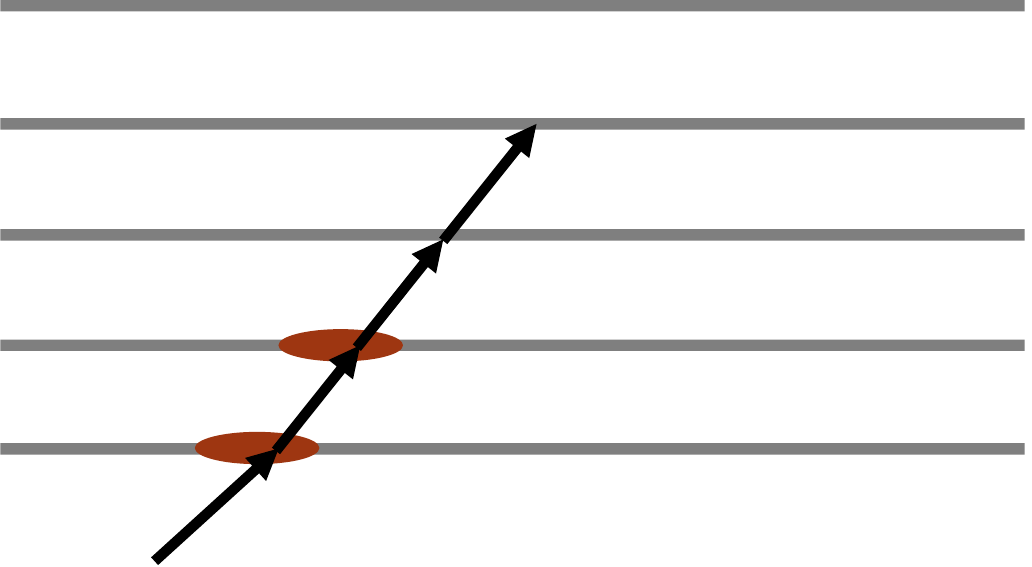}
    \caption{Illustration of longitudinal likelihood calculation. The gray horizontal lines 
    represent five KLM layers, with associated hits (brown ellipses) on the first two layers. 
    The track extrapolation represented by the black arrows stops at the fourth layer. In this case, 
    we have $N_{ext} = 4$ and 
    longitudinal likelihood of hypothesis $t$ given by $\mathcal{L}_t^{\rm long} = P_{t, 1}\varepsilon_1 P_{t, 2}\varepsilon_2 (1-P_{t, 3}\varepsilon_3)(1-P_{t, 4}\varepsilon_4)$. }
    \label{fig:NhitandNext}
\end{figure}

The longitudinal likelihood $\mathcal{L}^{\rm long}_t = \prod_{n=1}^{N_{ext}} \mathcal{L}_{t, n}$ is calculated from the likelihoods
\begin{equation}
\label{eq:longitudinal}
\mathcal{L}_{t, n}=\left\{\begin{array}{c}
P_{t, n}\cdot \varepsilon_n, \quad \text { with associated hit } \\
1-P_{t, n} \cdot \varepsilon_n, \quad \text { without associated hit }
\end{array}\right.
\end{equation}
of hit pattern in the layers crossed by the extrapolation track. 
$P_{t,n}$ stands for the probability that track $t$ penetrates to layer~$n$ 
as a function of extrapolated stopping layer $N_{ext}$. $P_{t,n}$ is measured in the simulation sample in advance. 
Detector efficiencies $\varepsilon_n$ are considered as well, and they are measured in data. 
In this work, detector efficiencies are assumed to be 100\%\footnote{Typically a 96\% hit detection efficiency is achieved in operation.}. 
An illustration of the longitudinal likelihood calculation is given in Fig.~\ref{fig:NhitandNext}. 

\begin{figure}[h]
    \centering
    \includegraphics[width = 0.48 \textwidth]{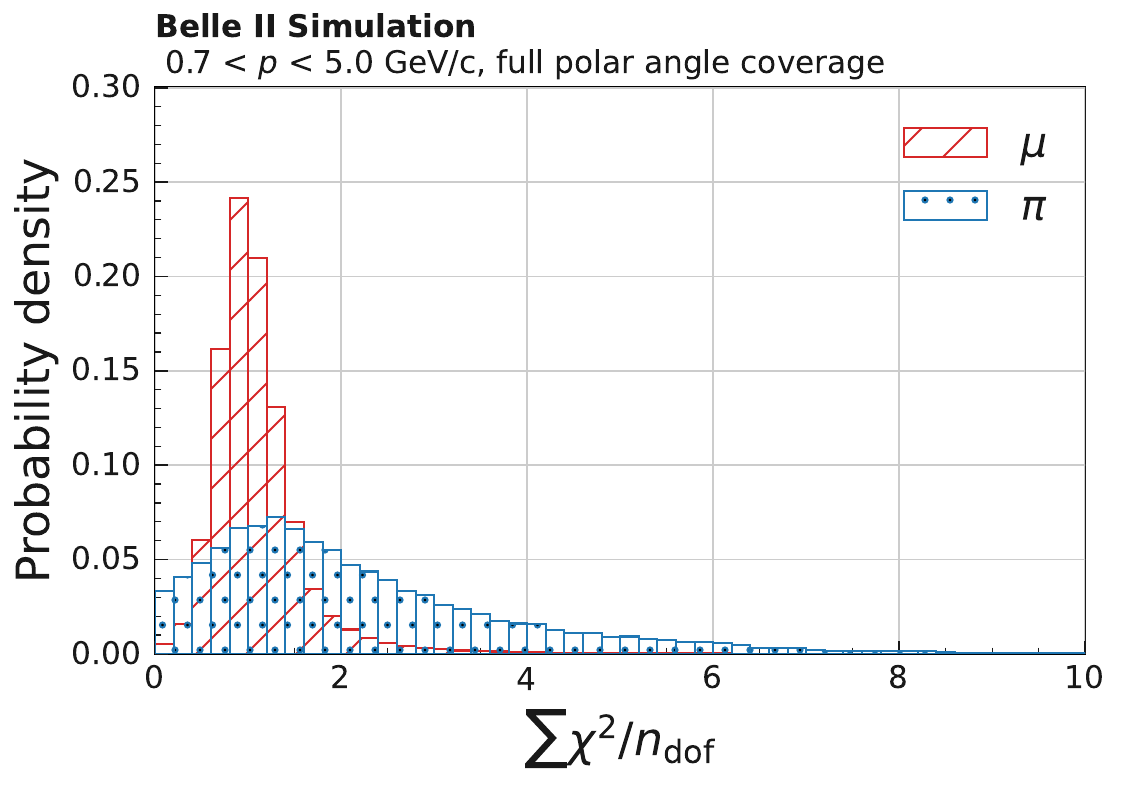}
    \caption{Distribution of $\sum\chi^2/n_{\rm dof}$ for muon (red, dashed) and pion (blue, dotted). }
    \label{fig:chi2distribution}
\end{figure}

The transverse likelihood is estimated based on the extrapolation quality described 
by the sum of $\chi^2$ of all associated hits ($\sum \chi^2$) and 
the number of degrees of freedom ($n_{\rm dof}$). The $n_{\rm dof}$ is twice the number of associated hits, because the $\chi^2$ of every hit is calculated once for each of the two directions. 
For muons, the distribution of $\sum\chi^2/n_{\rm dof}$ 
peaks at 1 while for pions, the distribution is wider due to multiple scattering 
inside KLM, 
as shown in Fig.~\ref{fig:chi2distribution}. 

\subsection{Discussion}

The muonID shows a good performance in $\mu/\pi$ separation. Still, some room for 
improvement is found, and will be discussed below. 

\begin{figure}[htpb]
    \centering
    \begin{subfigure}[b]{0.48\textwidth}
        \centering
        \includegraphics[width=\textwidth]{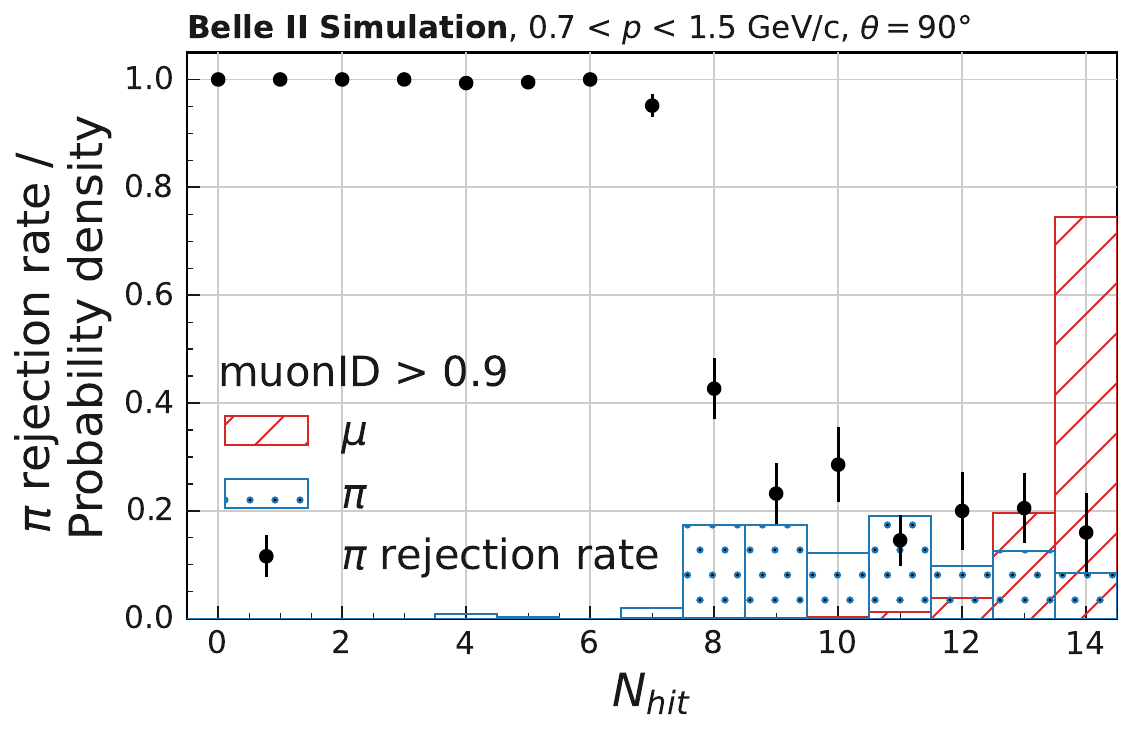}
        \caption{}
    \end{subfigure}
    \\
    \vspace{2mm}
    \begin{subfigure}[b]{0.48\textwidth}
        \centering
        \includegraphics[width =\textwidth]{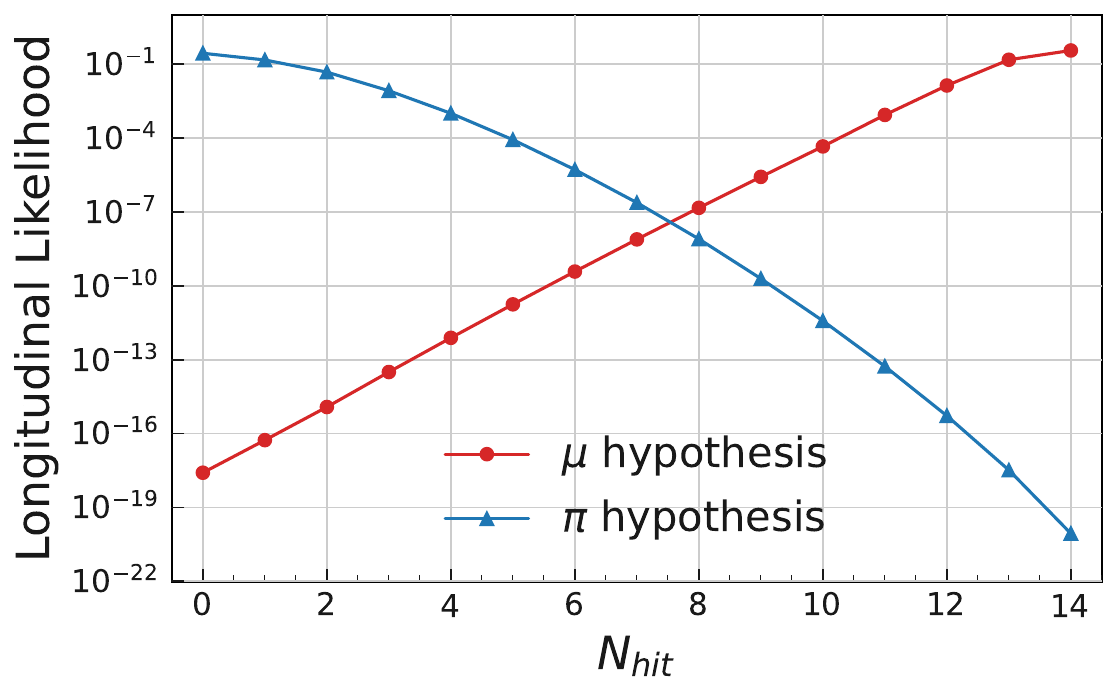}
        \caption{}
    \end{subfigure}
    \caption{(a): dots show the pion rejection rate against penetration layer after requiring 
    muonID $>$ 0.9 in samples with $N_{ext}=14$. Error bars represent statistical uncertainty. 
    The histograms are the 
    penetration layer distribution of remaining muon (red, dashed) and pion (blue, dotted)
    after selection. 
    (b): longitudinal likelihood (in logarithmic scale) under muon (red, circle) and pion (blue, triangle) hypothesis 
    as a function of penetration layer assuming extrapolation stops at layer 14 of BKLM. }
    \label{fig:possibleimprovement}
\end{figure}

\begin{figure}[htpb]
    \centering
    \includegraphics[width = 0.48 \textwidth]{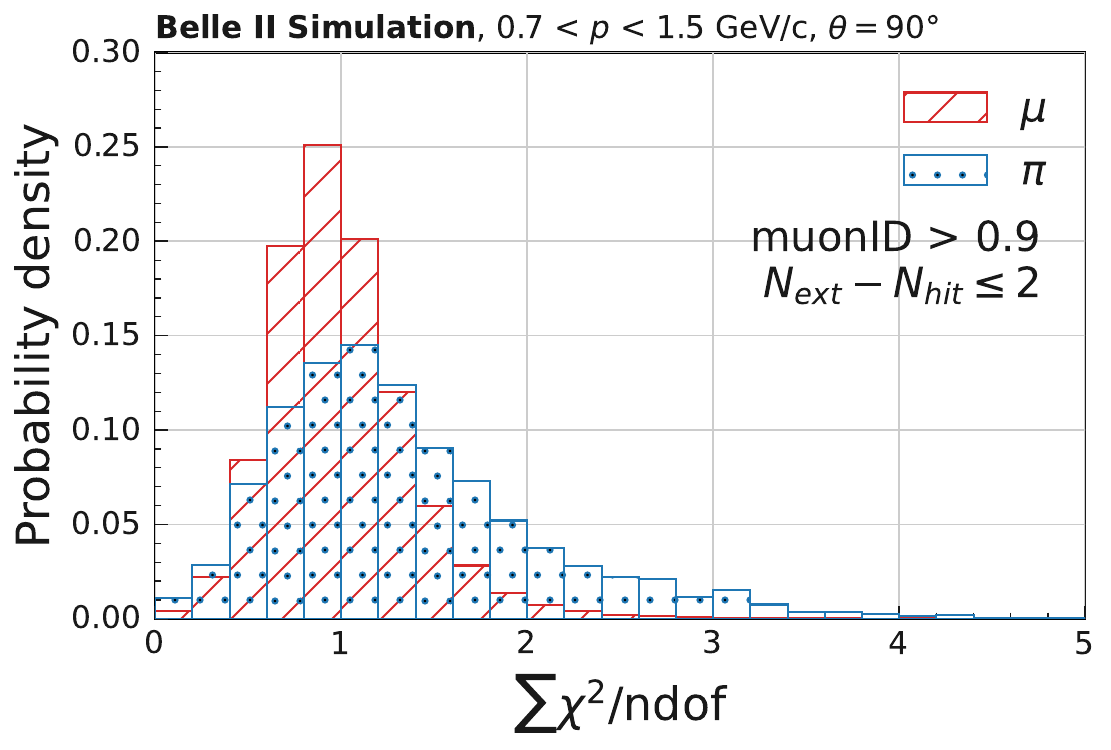}
    \caption{Distribution of $\sum\chi^2/n_{\rm dof}$ for muon (red, dashed) and pion (blue, dotted) samples, requiring muonID $>0.9$ and $N_{ext} - N_{hit} \leq 2$. This plot indicates that $\sum\chi^2/n_{\rm dof}$ information is not fully exploited by muonID. }
    \label{fig:transverse}
\end{figure}

Figure~\ref{fig:possibleimprovement}(a) shows the pion rejection rate 
as a function of the penetration layer ($N_{hit}$) applying a muonID $>$ 0.9 
selection, where $N_{hit}$ is defined as the last layer in which a hit has been detected. 
The histograms show the probability density of the penetration layer after 
selection for muon and pion samples. 
This dataset is a simulation sample with only one track in each event. 
For illustration, the polar angle is fixed to 90\degrees and only tracks 
extrapolated to stop at layer 14 of BKLM are selected. 
From this figure, it is obvious that muonID 
successfully rejected pions with penetration layers smaller than eight. However, 
the rejection rate reduces to only around 20\% when $N_{hit} > 8$. 
Meanwhile, we can significantly improve the identification performance by rejecting tracks in the range of $8\leq N_{hit} \leq 10$, 
which happens rarely for the muons, but quite frequently for the pions. 

To explore the reason why muonID failed to reject pion tracks satisfying $8\leq N_{hit} \leq 10$, 
we calculate the longitudinal likelihood $\mathcal{L}^{\rm long}$ as a function of
penetration layer ($N_{hit}$) with muon and pion hypotheses\footnote{In this calculation, we assume extrapolation stops 
at layer 14 of BKLM, associated hits observed up to layer~$N_{hit}$, and
no associated hit observed above layer~$N_{hit}$: $\mathcal{L}^{\rm long}_t = \prod_{n=1}^{N_{hit}} \varepsilon_n P_{t, n} \prod_{n=N_{hit}+1}^{14} (1-\varepsilon_n P_{t, n})$.}.
The result is presented in Fig.~\ref{fig:possibleimprovement}(b) and it shows that 
$\mathcal{L}^{\rm long}_{\mu}$ overwhelms $\mathcal{L}^{\rm long}_{\pi}$ when 
the track penetration layer is greater than 8, which explains the drastic drop in the rejection 
rate at layer 8.
It suggests that the longitudinal likelihood used in muonID is not optimally modeled, indicating significant potential for performance improvement. 
One possible explanation for this mis-modeling is that 
the longitudinal likelihood does not consider the correlations between hits in different layers, 
as it is constructed by simply multiplying the likelihoods assigned to individual layers.
This insight motivates us to develop a machine learning-based algorithm capable of incorporating such correlations in this work.

\begin{figure*}[htpb]
    \centering
    \begin{subfigure}[b]{0.4\textwidth}
        \centering
        \includegraphics[width=\textwidth]{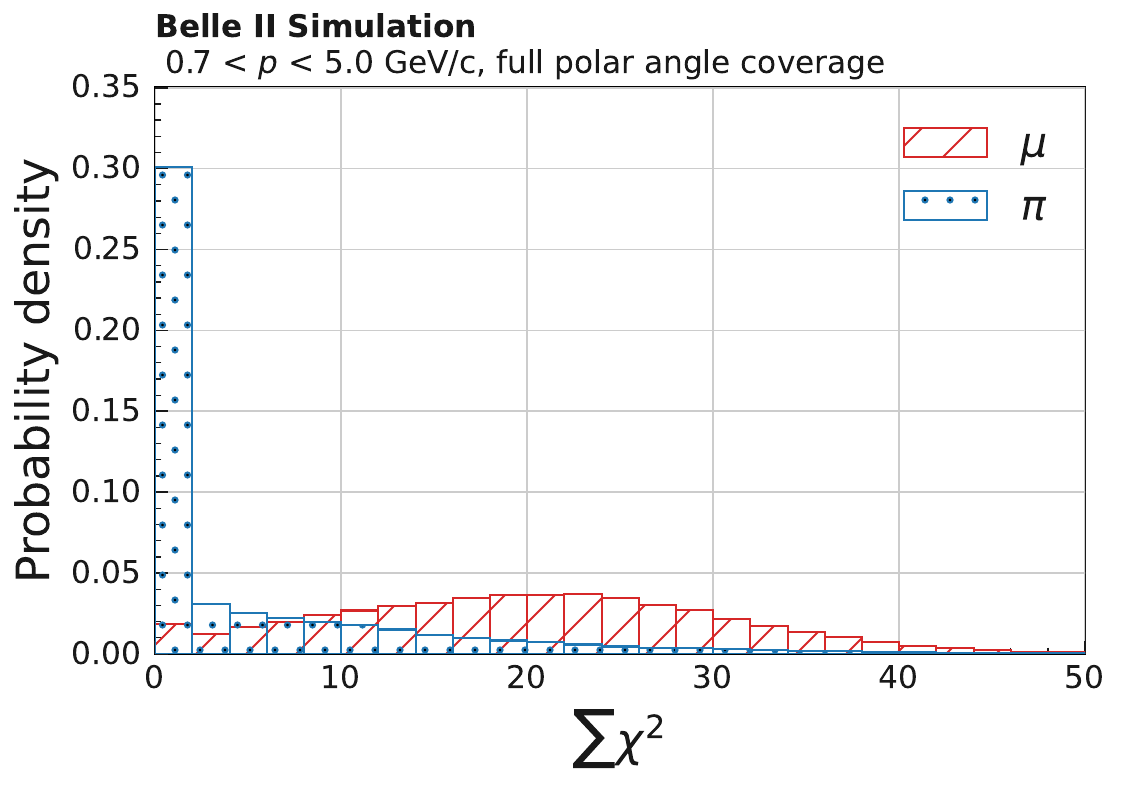}
        \caption{}
    \end{subfigure}
    \begin{subfigure}[b]{0.4\textwidth}
        \centering
        \includegraphics[width =\textwidth]{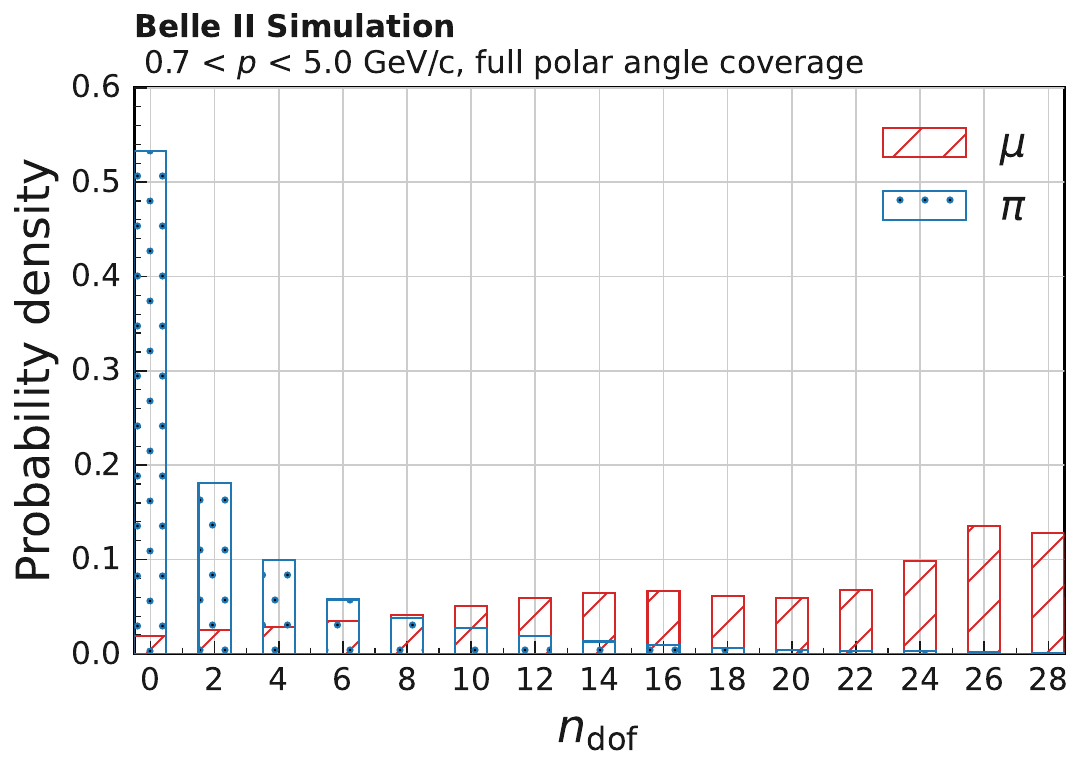}
        \caption{}
    \end{subfigure}
    \begin{subfigure}[b]{0.4\textwidth}
        \centering
        \includegraphics[width =\textwidth]{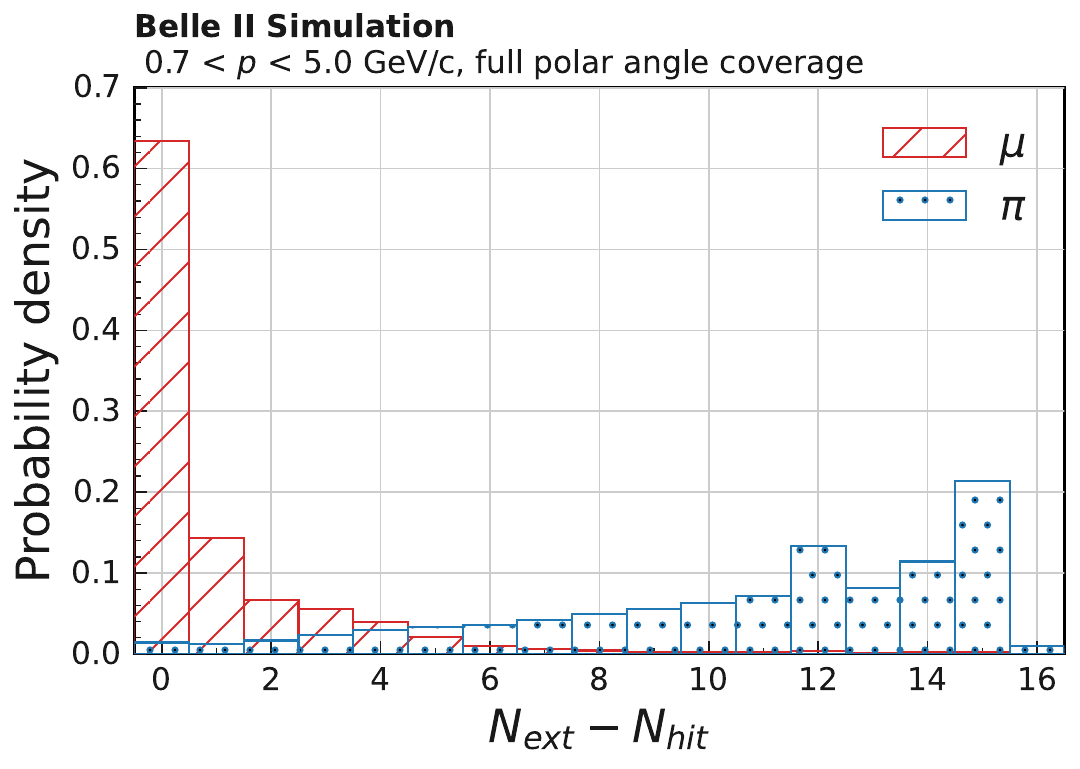}
        \caption{}
    \end{subfigure}
    \begin{subfigure}[b]{0.4\textwidth}
        \centering
        \includegraphics[width =\textwidth]{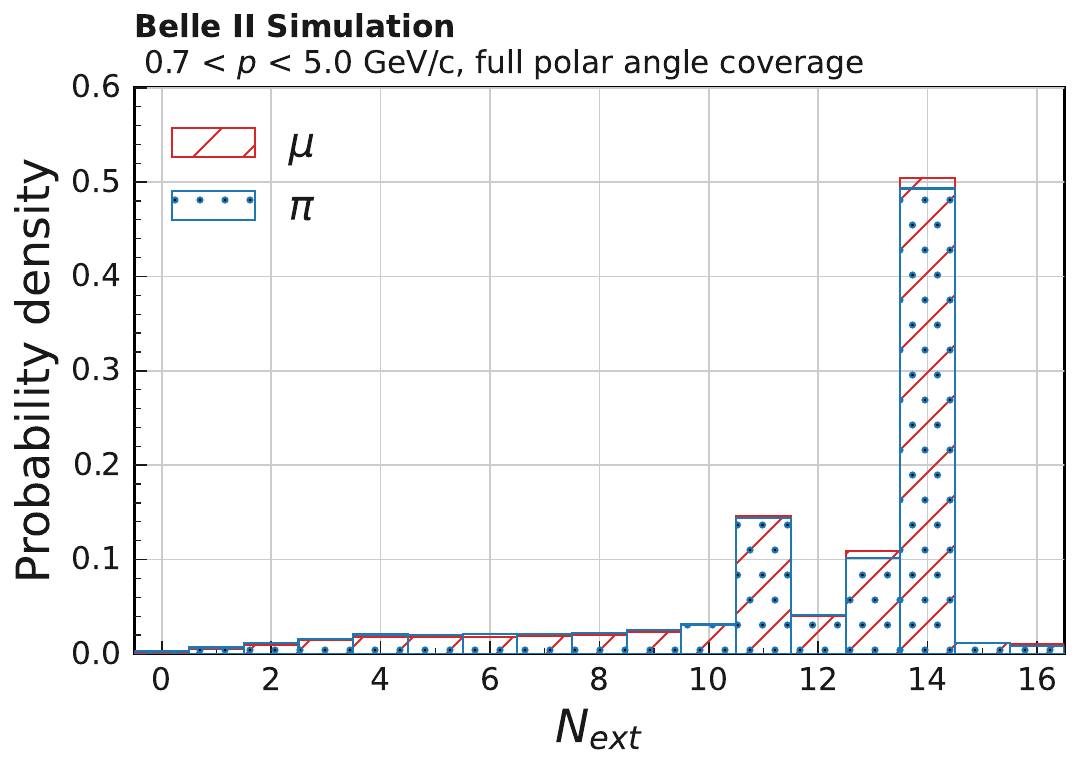}
        \caption{}
    \end{subfigure}
    \begin{subfigure}[b]{0.4\textwidth}
        \centering
        \includegraphics[width =\textwidth]{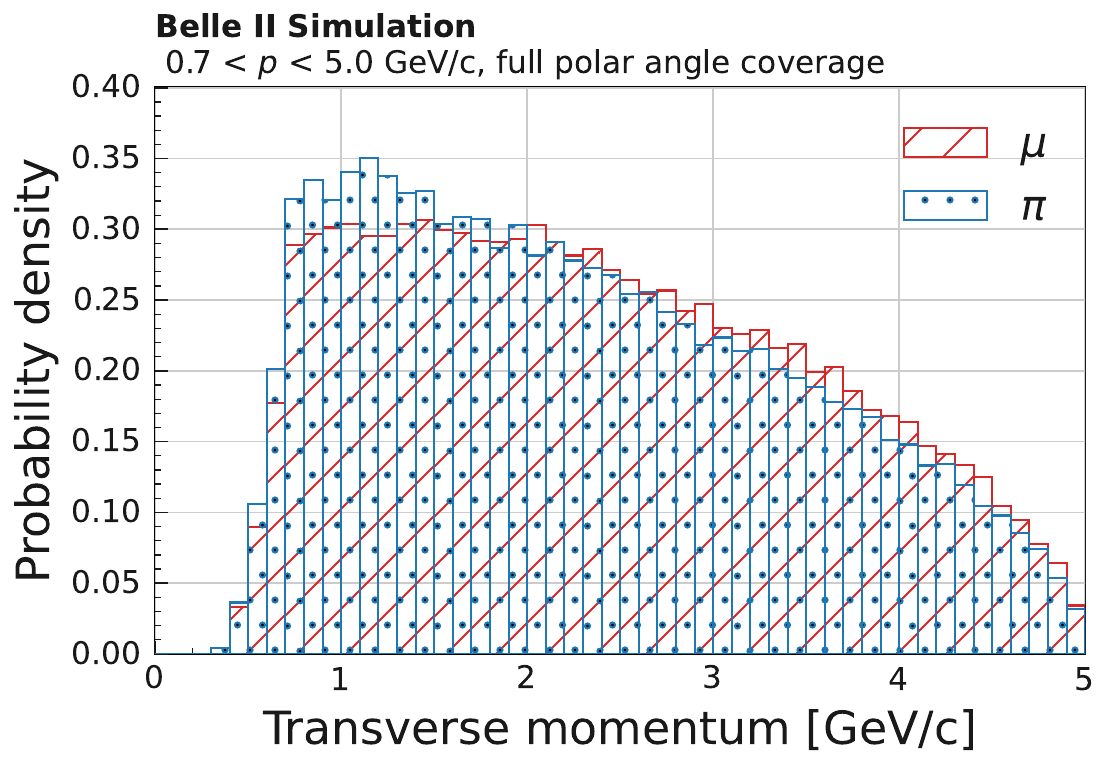}
        \caption{}
    \end{subfigure}
    \begin{subfigure}[b]{0.4\textwidth}
        \centering
        \includegraphics[width=\textwidth]{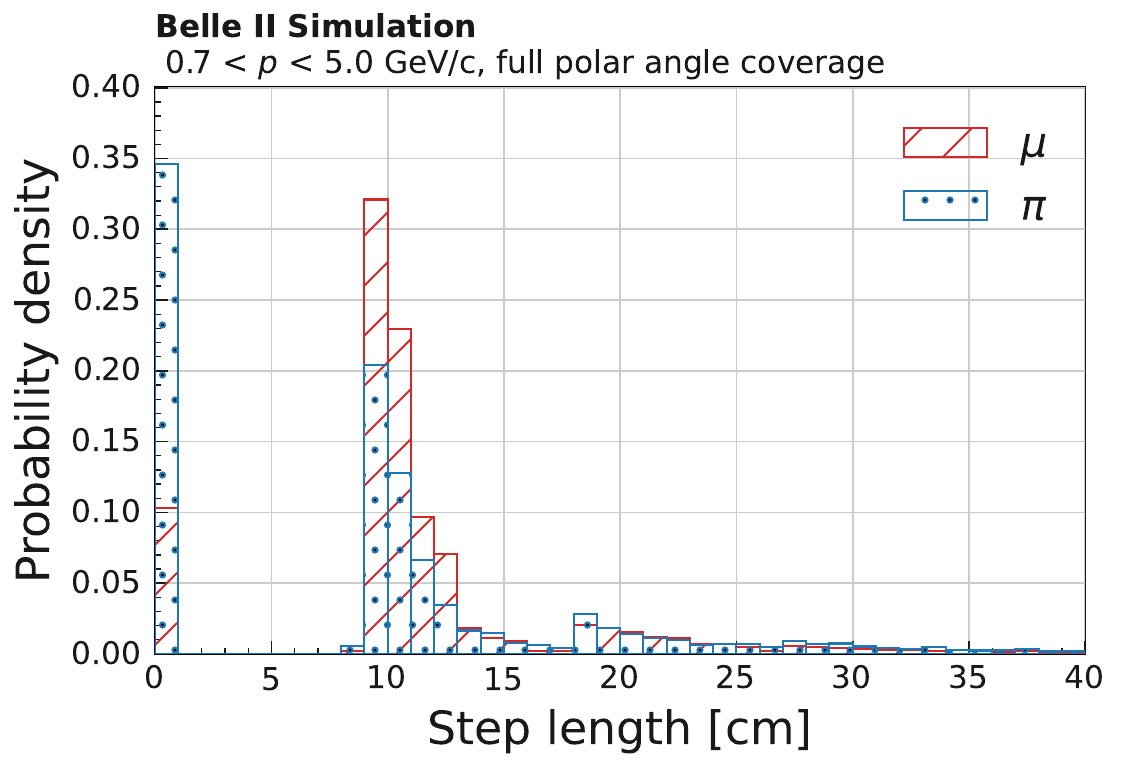}
        \caption{}
    \end{subfigure}
    \begin{subfigure}[b]{0.4\textwidth}
        \centering
        \includegraphics[width=\textwidth]{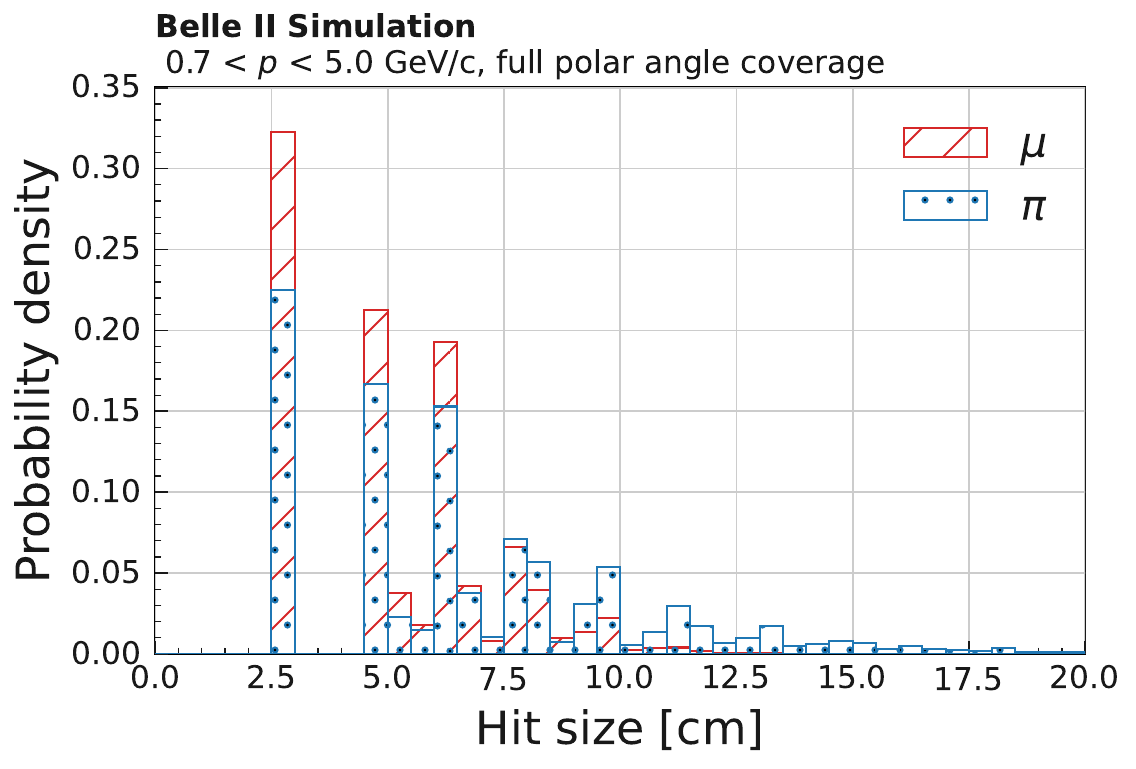}
        \caption{}
    \end{subfigure}
    \begin{subfigure}[b]{0.4\textwidth}
        \centering
        \includegraphics[width =\textwidth]{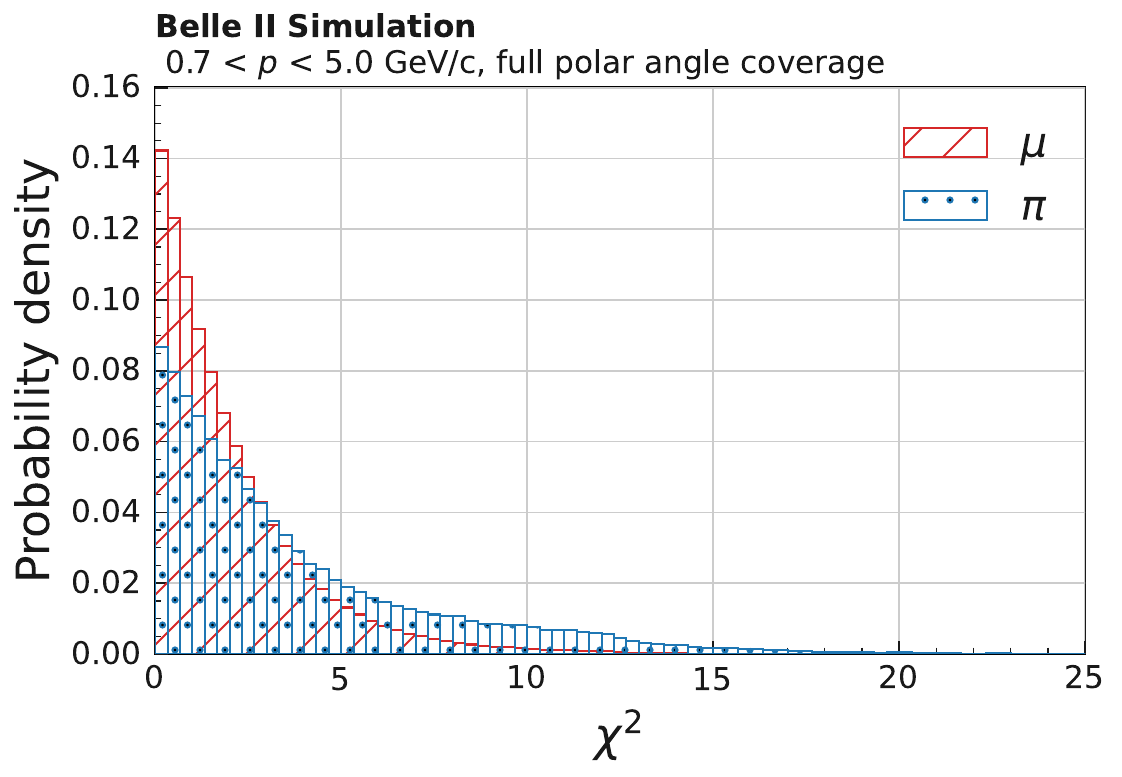}
        \caption{}
    \end{subfigure}
    \caption{Distributions of input variables used in DNN in the training sample. Plot (a)--(e) show the distributions of global variables with each entry in the histograms representing one track. Plot (f)--(h) show the distributions of hit pattern variables with each entry in the histograms representing one associated hit. The peak at zero in (f) represents the first associated hits of each track, whose step length are assigned to zero by definition. The peak around 2.5~cm in (g) represents hits with only one strip in both directions. All plots are normalized to unit area.}
    \label{fig:InputVariables}
\end{figure*}

In addition, there is room for improvement by better utilizing the transverse information.  
Fig.~\ref{fig:transverse} shows the 
distribution of $\sum\chi^2/n_{\rm dof}$ for tracks satisfying muonID $>0.9$ 
and $N_{ext} - N_{hit} \leq 2$. Still, some remaining pions can be rejected 
by requiring, for example, $\sum\chi^2/n_{\rm dof} < 2$, at the cost of losing little muon efficiency. 
However, muonID fails to do so because it relies too heavily on longitudinal 
information due to imperfect settings of the scale of 
longitudinal and transverse likelihood. 
Specifically, $\mathcal{L}^{\rm trans}_\mu/\mathcal{L}^{\rm trans}_\pi$ 
is at the order of $10^{-1}$ for tracks with $\sum\chi^2/n_{\rm dof} > 2$ , while 
$\mathcal{L}^{\rm long}_\mu/\mathcal{L}^{\rm long}_\pi$ is larger than $10^{-2}/10^{-15}=10^{13}$ 
according to Fig.~\ref{fig:possibleimprovement}(b). 
For this reason, the relationship muonID $>0.9$ 
is hardly influenced by the transverse likelihood.

\section{Deep Neural Network (DNN) based muon probability}

To make better use of penetration and transverse information, we 
propose a DNN-based algorithm. 
The track extrapolation and associated hits information described in Sec.~\ref{sec:Track extrapolation} are used in this algorithm.
In this section, input variables, network structure and training, as well as the 
performance evaluation of the new algorithm are described.

\subsection{Input variables}

Five global variables are used as input of the DNN, arranged in the order 
of $\sum\chi^2$, $n_{\rm dof}$, $N_{ext} - N_{hit}$, $N_{ext}$ and the transverse momentum 
of the track whose distributions are shown in Fig.~\ref{fig:InputVariables}(a)--(e). 
The latter two variables play an important role of indicator since 
the distributions of the former three variables as well as the hit pattern 
vary as function of the extrapolation layer and the transverse momentum. 

\begin{figure}[htpb]
    \centering
    \includegraphics[width = 0.48 \textwidth]{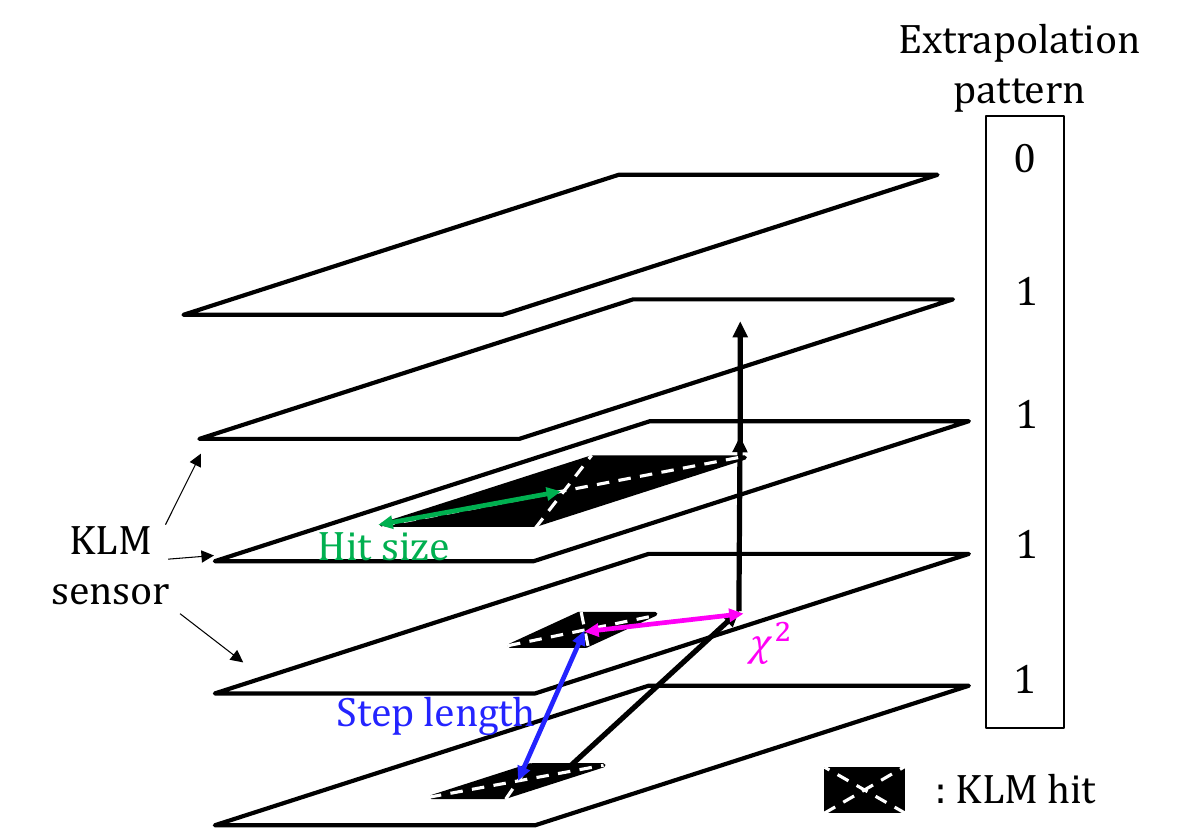}
    \caption{Illustration of hit pattern variables. The thick black arrows represents the track extrapolation and the extrapolated position at the third layer is adjusted by Kalman-filter. 
    The length of the blue and green arrows represents step length and hit size, respectively. 
    The magenta arrow indicates the $\chi^2$ between hit and extrapolation position. 
    The binary numbers on the right side are the extrapolation pattern of the corresponding layers. }
    \label{fig:hitpattern}
\end{figure}

In addition, four hit pattern variables are defined for each KLM layer 
used as input of the DNN as illustrated in Fig.~\ref{fig:hitpattern}. 
Their definitions are explained below. 

\textit{Step length:} defined for each associated hit as the 
distance to its prior associated hit. If it happens to be the 
first associated hit of the track, its step length is 
set to zero. 

\textit{Hit size:} defined for each associated hit as being 
half of the diagonal length of the rectangular shape of the hit.

\textit{$\chi^2$:} defined in Eq.~\ref{eq:chi2}.

\textit{Extrapolation pattern:} a binary value indicating whether 
the extrapolation crossed the corresponding layer or not.

The calculation of muonID longitudinal likelihood in Eq.~\ref{eq:longitudinal} is layer-based, which means that it only reflects the penetration information along the normal direction of the detector layer plane. 
By introducing the step length into the DNN, 
the penetration information along the tangent direction (projection of track direction on the sensor plane) is also taken into account.
Due to the stronger penetration ability, the total 
penetration depth (sum of step length) of the muon is greater than that of the pion.
And on the other hand, 
the variation of step length between different layers of pions tends 
to be larger than that of muons because of strong interaction 
with detector materials. 
For the same reason, the hit size and the 
deviation of extrapolation to the hit position ($\chi^2$) of the pion also tends to be 
larger than that of the muon, as shown in Fig.~\ref{fig:InputVariables}(f)--(h). 

In total, the input to the DNN model is a 1-dimensional float array with 121 elements. 
The first five elements 
are the global variables. 
The remaining 116 elements are arranged into 29 groups, each group is used to place the 
hit pattern variables of one layer. The first 15 groups represent the 15 BKLM layers,  
while the latter 14 groups represent the 14 EKLM layers. In each group, the hit pattern 
variables are arranged in the order of hit size, step length, $\chi^2$ and  
extrapolation pattern. If there is no associated hit in the corresponding layer, 
the hit size, step length, and $\chi^2$ are set to -1. 

\subsection{Network structure and training}

\begin{figure}[h]
    \centering
    \includegraphics[width = 0.48 \textwidth]{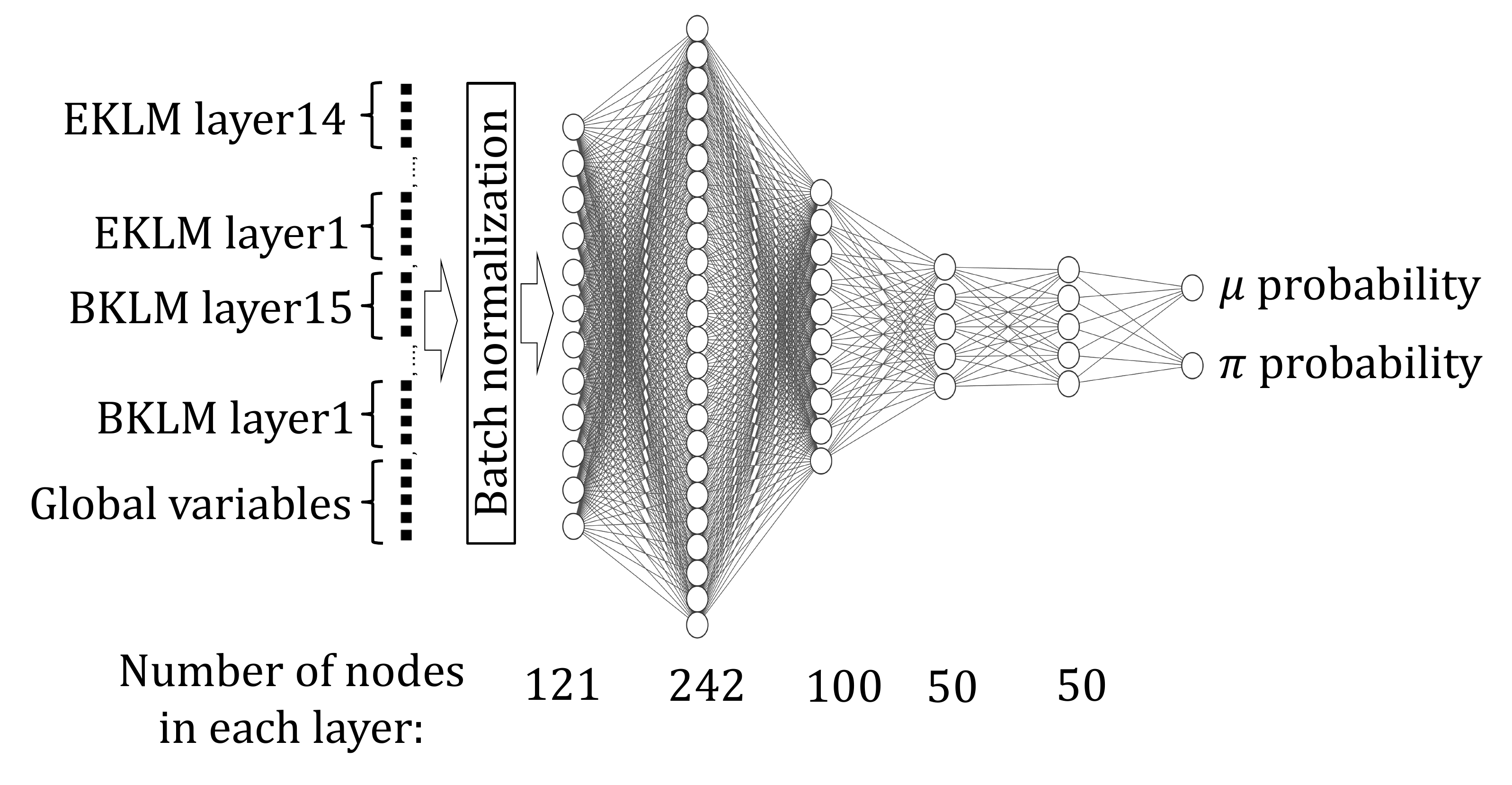}
    \caption{Structure of DNN. }
    \label{fig:NNstructure}
\end{figure}

This neural network is built with the PyTorch~\cite{pytorch} library 
and its structure is shown in Fig.~\ref{fig:NNstructure}. 
The input array is first processed 
by the batch normalization module, followed by a fully connected 
linear model. There are five linear layers with 121, 242, 100, 50 and 50 nodes, respectively.
The output of each node is processed 
by a LeakyReLU~\cite{leakyrelu} activation function before being input to the next layer. 
At the output of the last layer there is a softmax 
activation function used to output the muon 
probability and the pion probability. In total, there are 64318 trainable parameters 
in the model. 

A simulation sample is generated for training, validation, and test 
of the model 
using the Belle~II Analysis Software Framework~\cite{basf2, basf2_code}. 
Each event contains 4 to 16 tracks to simulate different event multiplicity. 
Each track is randomly generated to be a muon, pion, electron, kaon, proton, or deuteron, with the same probability for each type. 
The charge of each track is also randomly determined to be positive or negative with equal probability. 
To improve the robustness against noise hits, simulated beam background~\cite{beambkg} at a luminosity about six times higher than the current operation record is overlaid on each event. 
All tracks are generated with uniform momentum ranging from 0.7~\gevc to 5.0~\gevc, 
cosine of polar angle and azimuthal angle distribution, 
covering the full geometric acceptance of KLM. 
Only the muons and pions in the samples are selected for study using generator information.  
In addition, pions that decay before KLM are removed from the samples. 
In total, we generated 559383, 338031 and 153966 tracks for training, validation, 
and test samples, respectively. 

\begin{figure}[h]
    \centering
    \begin{subfigure}[b]{0.45\textwidth}
        \centering
        \includegraphics[width=\textwidth]{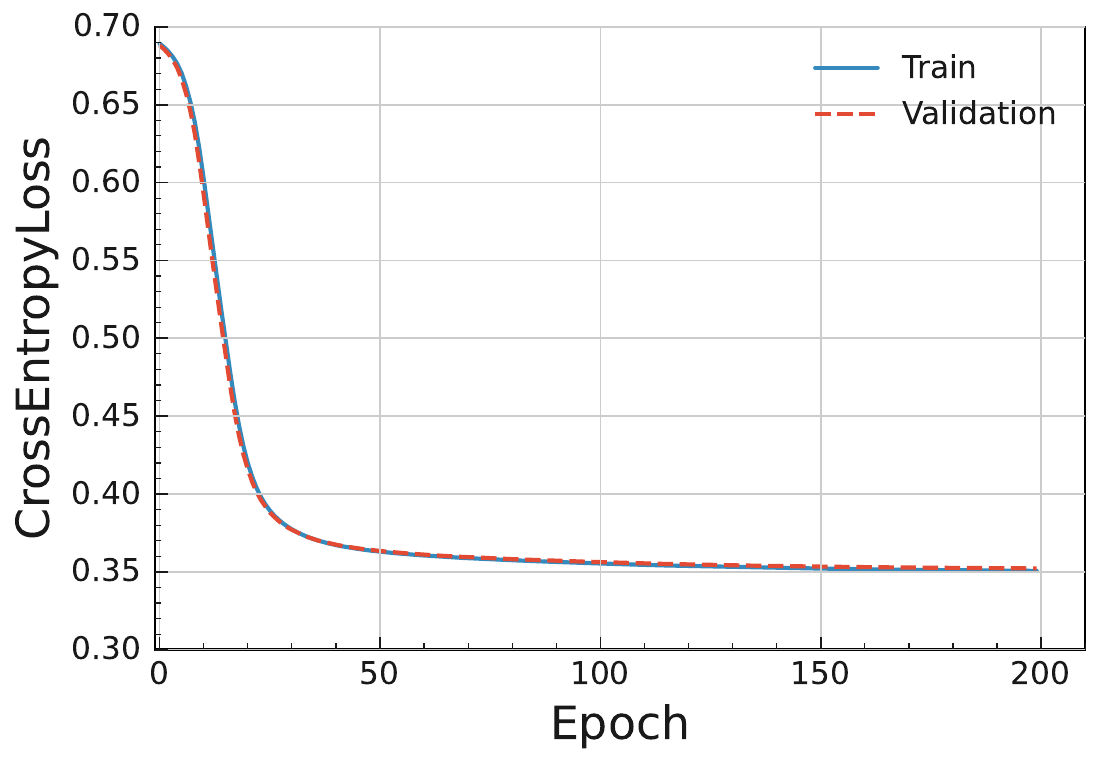}
        \caption{}
    \end{subfigure}
    \\
    \vspace{2mm}
    \begin{subfigure}[b]{0.45\textwidth}
        \centering
        \includegraphics[width =\textwidth]{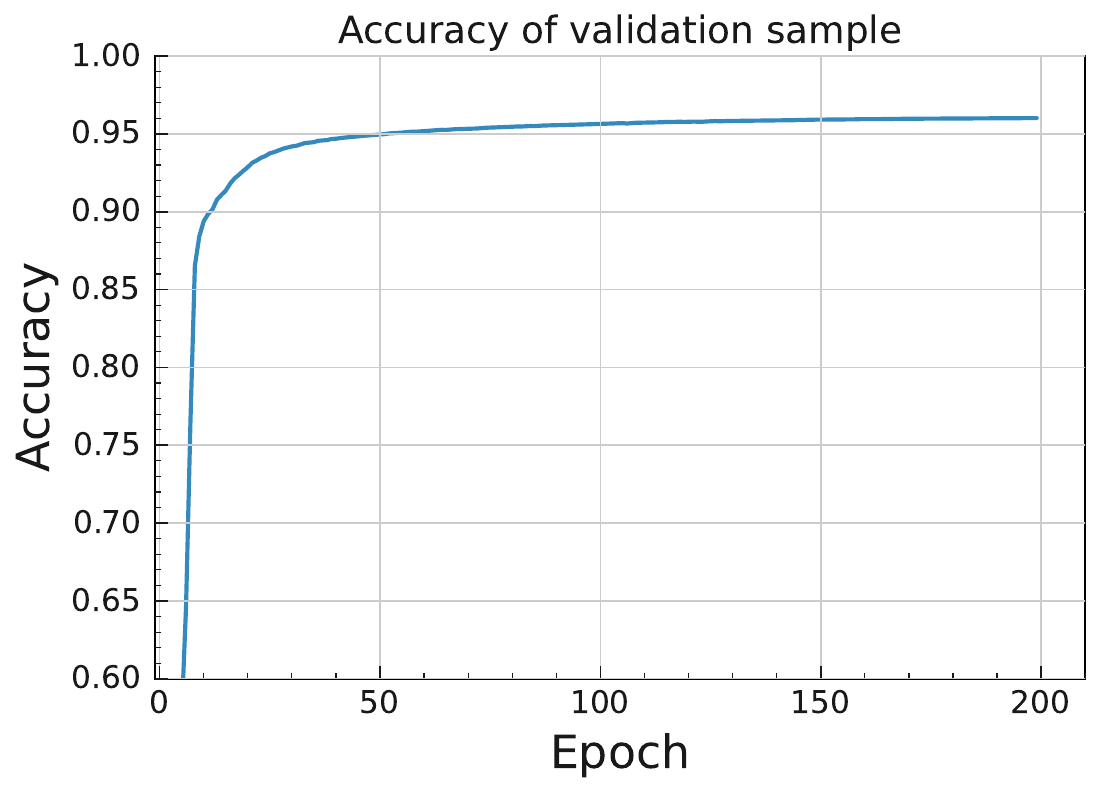}
        \caption{}
    \end{subfigure}
    \caption{(a): Loss convergence plot of training (blue, solid) and validation (red, dashed) samples. (b): Accuracy of validation sample as a function of epochs. }
    \label{fig:training_data}
\end{figure}

In the training, an \texttt{Adam} optimizer is adapted with a learning rate
of $10^{-5}$ and the batch size is set to be 10000.  
\texttt{CrossEntropy}, which is suitable for binary classification, is used as the loss to minimize. 
Training stops when prediction accuracy of the validation sample does not increase for 10 epochs and the epoch with the best accuracy is adapted. 
The loss convergence plots of training and validation samples, as well as the prediction accuracy in validation sample as a function of epochs are presented in Fig.~\ref{fig:training_data}. 
The muon efficiency difference at 2\% pion fake rate in training, validation and test samples are at the order of $O(0.1\%)$, indicating that no significant over-training is observed. 
The training is performed on one GeForce RTX 3090 GPU with a training time of around 30~min. 
Inference latency is $1.81\pm0.78$~ms per $e^+e^-\to\Upsilon(4S)\to B\bar{B}$ event on the KEKCC CPU cluster (AMD EPYC 9534) at KEK. 

\subsection{Performance}
\label{sec:Performance}

\begin{figure}[h]
    \centering
    \includegraphics[width = 0.48 \textwidth]{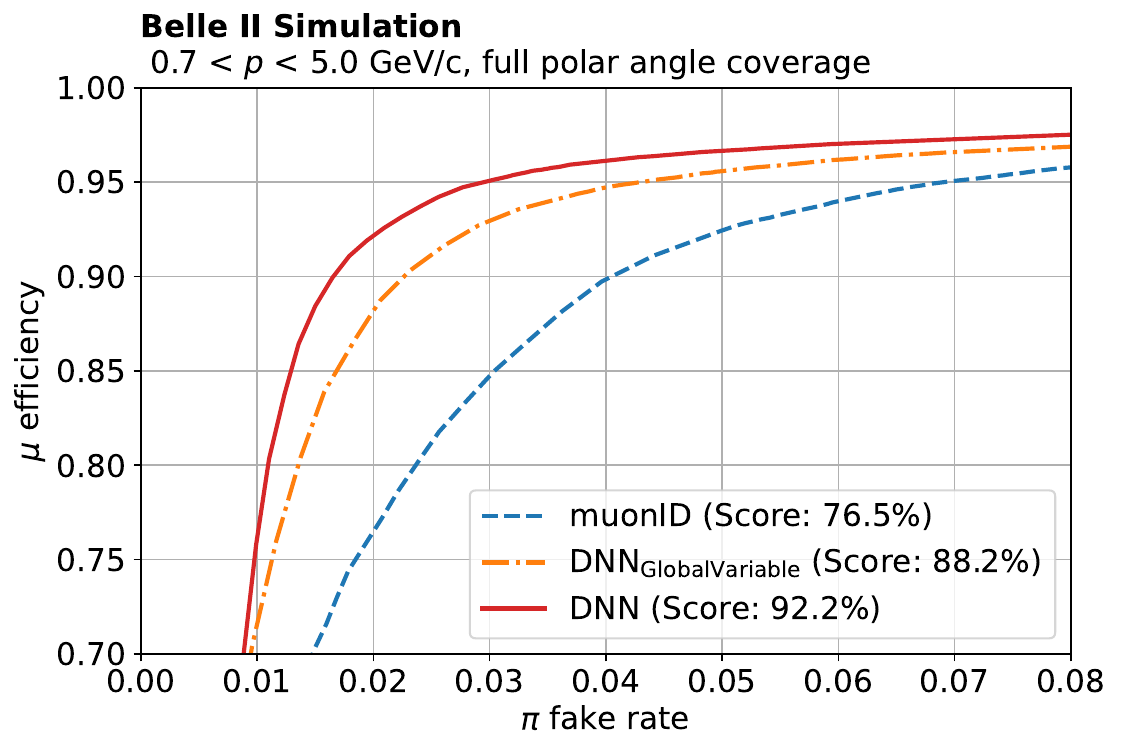}
    \caption{ROC curve of muonID (blue, dashed), DNN trained with only global variables (orange, dashdot), and the default DNN (red, solid). The score of each model is defined as the muon efficiency at 2\% pion fake rate in the test sample. }
    \label{fig:roccurve}
\end{figure}

The performance of the model is validated using a Receiver Operation 
Characteristic (ROC) curve, which plots the true positive rate ($\mu$ efficiency) 
against the false positive rate ($\pi$ fake rate) as shown in Fig.~\ref{fig:roccurve}. 
The muonID, which is the baseline method, is also plotted for comparison. 
As demonstrated in the ROC curves, the DNN performs better than muonID. 
For example, the DNN (muonID) gives a pion fake rate of 1.6\% (4.1\%) at 
90\% muon efficiency, or a muon efficiency of 92.2\% (76.5\%) at a pion fake rate of 2\%. 
To validate the importance of the hit pattern variables, we trained another network using only the five global variables as input.
The structure of the network is identical to the default one, except for the batch normalization and the first linear layer, whose number of nodes are adjusted according to input array length. 
The pion fake rate deteriorates to 2.3\% at 90\% muon efficiency if we only use the 
five global variables, demonstrating the importance of the hit pattern.

To test the robustness of the model against background hit, we prepared another sample which is generated in the same way as training sample, except that the beam induced background simulated at the Belle~II nominal luminosity, which is about three times larger than the training sample. 
In this sample, the pion fake rate deteriorates to 2.4\% at 90\% muon efficiency possibly due to the changes of hit patterns under high background environment, but still it performs significantly better than muonID. 
A re-training of the DNN model can be expected when machine status evolves significantly to achieve a better performance. 
We also tested the model performance by setting the KLM hit detection efficiency to a uniform 85\% and observed slight pion fake rate increased to 2.0\%. 
Finally, we verify the magnetic field inside KLM by 10\% to test its robustness against inaccuracies in track extrapolation and no significant performance variance is observed. 

\begin{figure}[htpb]
    \centering
    \includegraphics[width = 0.48 \textwidth]{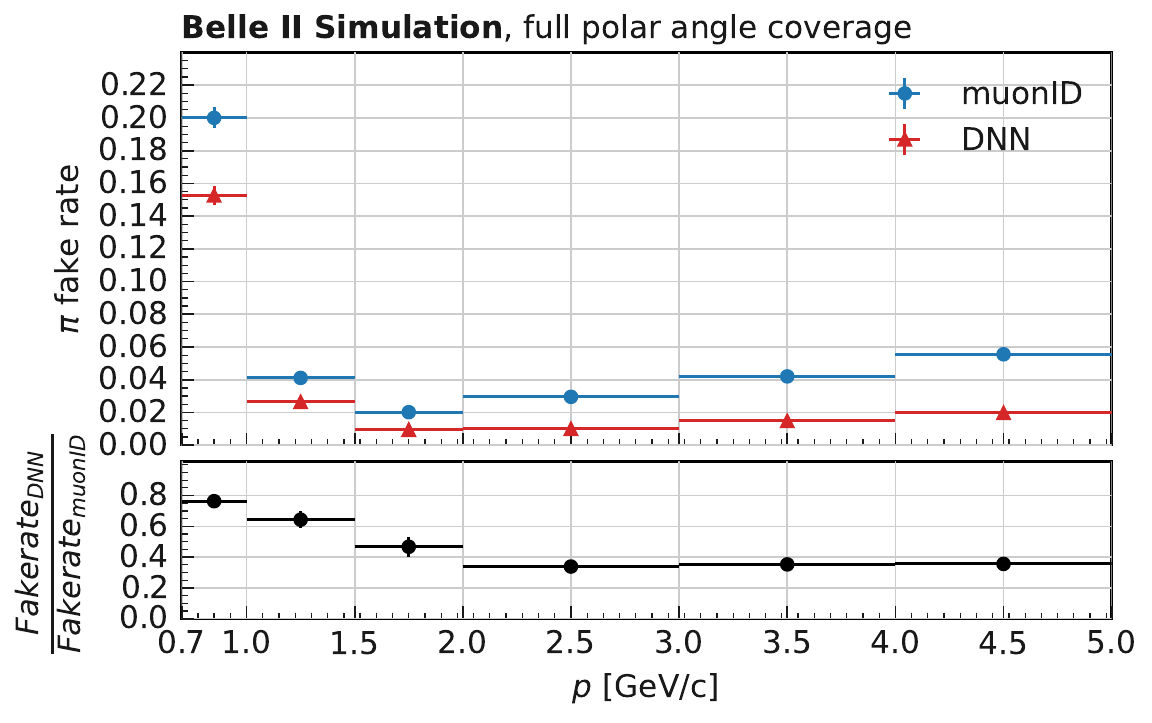}
    \caption{Upper: pion fake rate of muonID and DNN as a function of track momentum at 90\% uniform muon efficiency. Lower: pion fake rate  ratio of DNN over muonID. }
    \label{fig:MCperformance_pbin}
\end{figure}

Figure~\ref{fig:MCperformance_pbin} shows the pion fake rate of 
muonID and DNN at each momentum interval, maintaining a uniform muon efficiency of 90\%. 
The pion fake rate is suppressed across the full momentum range, with improvements exceeding 60\% in the high momentum range ($p>2.0~\gevc$). 
The improvement is less significant in the low-momentum region, where muons cannot traverse the KLM entirely. 

\begin{figure}[h]
    \centering
    \includegraphics[width = 0.48 \textwidth]{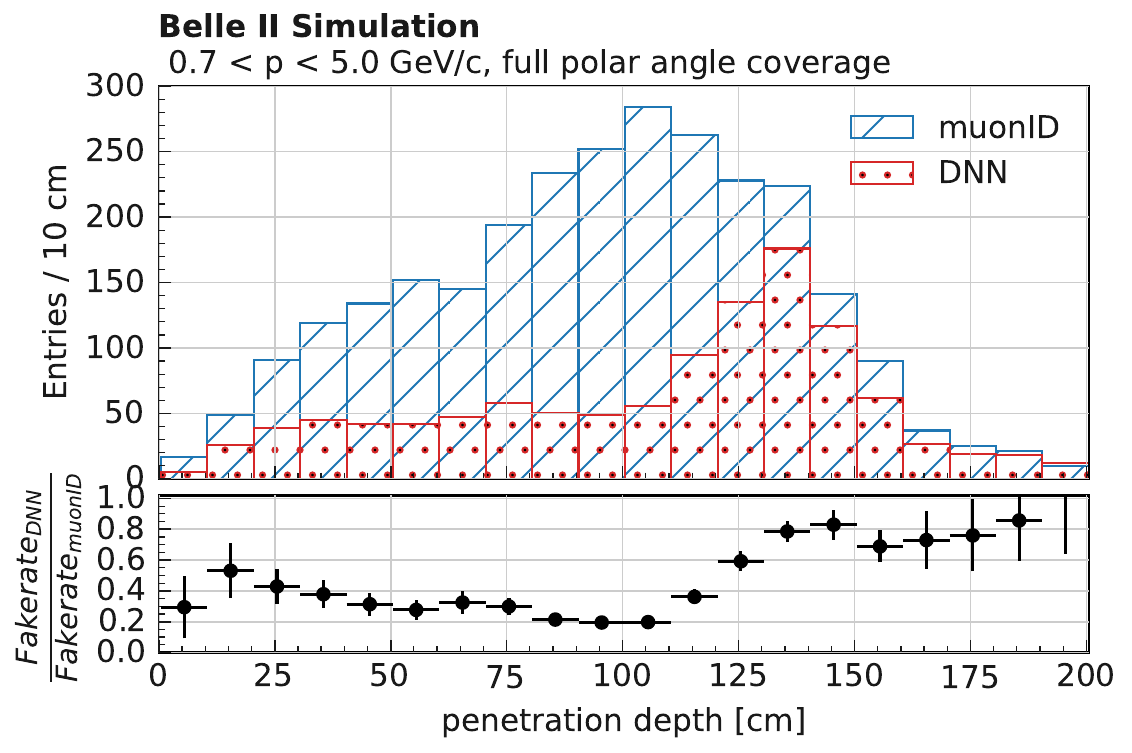}
    \caption{Upper: distribution of penetration depth of pion after the selection with 90\% overall muon efficiency for muonID and DNN method. Lower: pion fake rate ratio of DNN over muonID as a function of penetration depth.}
    \label{fig:penetrationdepth}
\end{figure}

Figure~\ref{fig:penetrationdepth}
shows the penetration depth distributions of the pions after selection 
at the overall muon efficiency 90\% using the muonID and DNN method, respectively.  
Comparing to muonID, DNN successfully rejected about 60\% of deeply penetrated pions up to a penetration 
depth of 125~cm, which aligns well with the detector thickness of around 130~cm for both BKLM and EKLM. 
This phenomenon may suggest that the DNN has learned a specific pattern: tracks with a penetration depth exceeding 125~cm are more likely to escape from the KLM, where $\mu/\pi$ identification based on penetration ability becomes less effective.

\section{Conclusion and prospects}
\label{}

In this paper, we discuss how the muon identification performance of the Belle~II experiment can be improved by better utilizing hit pattern information in the KLM detector. 
By training a new deep neural network, we reduced the pion fake 
rate (specificity) from 4.1\% to 1.6\% at 90\% muon efficiency (recall) in the simulation sample. 
This result is promising and this DNN has been implemented into the Belle~II Analysis Software Framework. 
A test of performance on real data is expected in the future. 

Further performance improvements are anticipated by integrating not only information from the KLM detector, but also combining output from the inner detectors.

\appendix

\section*{Author contribution statements}

Z.~Wang conceived of the presented idea, developed the code and wrote the manuscript under the supervision of Y.~Sato, A.~Ishikawa, Y.~Ushiroda, K.~Uno, and K.~Sumisawa. 
G.~De~Pietro and F.~Meier helped in coding and implementation into Belle II Analysis Software Framework. 
N.~K.~Baghel, S.~Choudhury, C.~Ketter, H.~Kindo, T.~Lam, and S.~Prell contributed to development, operation of KLM detector and data taking. 
All authors provided critical feedback to the manuscript. 

\section*{Acknowledgments}
This work is supported by the JSPS KAKENHI Grant Number JP24KJ0650 and JP22H00144. 






\end{document}